\documentclass[twocolumn,aps,prb,showpacs]{revtex4}
\usepackage{graphicx}
\usepackage{times}

\begin{document}

\title{Simulating a two-dimensional frustrated spin system with fermionic resonating-valence-bond states}

\author{Chung-Pin Chou$^{1}$}\author{Hong-Yi Chen$^{2}$}
\affiliation{$^{1}$Beijing Computational Science Research
Center,Beijing 100084, China} \affiliation{$^{2}$Department of
Physics, National Taiwan Normal University, Taipei 116, Taiwan}

\begin{abstract}
The frustrated Heisenberg $J_{1}-J_{2}$ model on a square lattice is
numerically investigated by variational Monte Carlo simulations. We
propose a antiferromagnetic fermion resonating-valence-bond
(AF-fRVB) state that has ability to examine the entire phase diagram
in the $J_{1}-J_{2}$ model. Two phase transition points, the second
order around $J_{2}/J_{1}=0.45$ and the first order around
$J_{2}/J_{1}=0.6$, can be extracted more clearly than the
conventional bosonic RVB state. At the maximally frustrated point
($J_{2}/J_{1}=0.5$), the AF-fRVB state shows the variational
ground-state energy in the thermodynamic limit very close to the one
estimated by the projected entangled pair state at the largest bond
dimension available. On the other hand, in the frustrated regime
$0.4\lesssim J_{2}/J_{1}\leq0.5$, AF-fRVB states with $s_{+-}$
(using the terminology in the field of iron-based superconductors)
and $d_{xy}$ pairing symmetries are degenerate in the thermodynamic
limit, implying the existence of gapless Dirac excitations in the
spinon spectrum.
\end{abstract}


\maketitle

\textit{Introduction.} Frustration is one of the simplest concepts
to induce a quantum phase transition in magnetic systems. Quantum
spin liquids, searched for both theoretically and experimentally
over decades, could be one of the products in frustrated spin models
\cite{BalentsNatPhys10,NormandCP09}. Notably, studies of quantum
phase transitions between spin-liquid phases and adjacent
magnetically ordered phases are important to understand quantum spin
liquids. To tackle the problem about the quantum phase transition, a
systematic analysis of the detailed ground-state phase diagram of
frustrated spin systems is required. The zero-temperature phase
diagram of the spin-1/2 $J_{1}-J_{2}$ square lattice model has been
reported by exact diagonalization (ED) calculations
\cite{MambriniPRB06,RichterEPJB10} and large-scale density matrix
renormalization group (DMRG) studies \cite{JiangPRB12,GongArXiv13}.
It is well known that the ground state displays a checkerboard
antiferromagnetic (AF) order at small $J_{2}/J_{1}$ and a collinear
AF order at large $J_{2}/J_{1}$. However, the existence of a gapless
or gapful quantum spin liquid between checkerboard and collinear AF
ordered phases has still remained unsolved.

To date, most of variational Monte Carlo (VMC) studies of the
$J_{1}-J_{2}$ model mainly focus on the maximally frustrated regime
($J_{2}/J_{1}\sim0.5$) and search a possible quantum spin liquid by
using either Schwinger bosonic or fermionic resonating-valence-bond
(RVB) wave functions
\cite{CapriottiPRL01,LouPRB07,MezzacapoPRB12,LiPRB12,WangPRL13,HuPRB13,QiArXiv}.
The RVB theory is the first proposal by Anderson to describe the
quantum spin liquid in a two-dimensional (2D) spin-1/2 Heisenberg
model \cite{AndersonSci87}. Aftermentioned, the bosonic RVB wave
function, categorized by the projective symmetry group
\cite{WenPRB02}, has been widely used to study different spin models
\cite{AuerbachPRL88,ArovasPRB88,FlintPRB09}. On the other hand, the
fermionic RVB wave function, constructed by the Gutzwiller
projection onto BCS mean-field states, has predicted the existence
of a gapless spin liquid in several different lattice structures
\cite{YunokiPRB06,RanPRL07,LiArXiv11}. However, both bosonic and
fermionic RVB states fail to demonstrate the quantum phase
transition involving the long-range magnetic order.

Recently, a tremendous numerical effort using the projected
entangled pair states (PEPS) has been performed. The numerical
result shows some missing data in a large part of the collinear regime
\cite{WangArXiv}. It can be expected that when the ground state is
on the verge of various instabilities around critical points
\cite{DarradiPRB08}, it is very difficult to distinguish the PEPS
with similar energies but different physical properties. On the
other hand, since the Gutzwiller projection enables the ground state
to recover symmetries lost in the BCS Hamiltonian, a
Gutzwiller-projected BCS wave function is invariant with respect to
the $SU(2)$ transformation implying high degeneracies after the
projection. An ideal Gutzwiller-projected wave function for the 2D
frustrated Heisenberg model can be thus obtained by using gap
functions with different pairing symmetries \cite{CapriottiPRB03}.

In this work, we simply extend the Gutzwiller-projected BCS wave
function to construct the fermionic RVB state which has explicit
AF magnetic orders, e.g. checkerboard or collinear long-range
patterns. We call it the AF fermion RVB (AF-fRVB) state. The
variational framework can demonstrate the phase transition between the
magnetic order and quantum spin disorder. Thus, this idea allows us
to determine the ground-state phase diagram of the $J_{1}-J_{2}$
model by using the VMC technique. Our main findings are the following:
(1) the zero-temperature phase diagram is successfully reproduced by
the AF-fRVB wave function; (2) a continuous phase transition near
$J_{2}/J_{1}\sim0.45$ and clear first-order phase transition at
$J_{2}/J_{1}=0.6$ are numerically confirmed; (3) a much less
computational cost in the AF-fRVB wave function than the PEPS is
performed. In particular, at $J_{2}/J_{1}=0.5$, the best energy
obtained from the AF-fRVB state is very close to the one reached by
the PEPS with rather large bond dimension; (4) in the highly
frustrated regime, the next-nearest-neighbor pairing symmetry of the
AF-fRVB state can be either $d_{xy}$ or $s^{+-}$. The $SU(2)$
symmetry suggests that the BCS Hamiltonian with Dirac nodes reflects
the gapless nature of the physical excitation spectrum.

\textit{Numerical Method.} We begin with the Hamiltonian,
\begin{eqnarray}
H=J_{1}\sum_{<i,j>}\mathbf{S}_{i}\cdot\mathbf{S}_{j}+J_{2}\sum_{\ll
i,j\gg}\mathbf{S}_{i}\cdot\mathbf{S}_{j},\label{e:eq01}
\end{eqnarray}
where $<i,j>$ and $\ll i,j\gg$ denote nearest and next-nearest
neighbors, respectively. $\mathbf{S}_{i}$ is the spin operator at
site $i$, and $J_{1}\equiv1,J_{2}>0$. We consider the $L\times L$
square lattice with periodic boundary condition of size
$L=8,16,20,24$. The AF-fRVB wave function we used here is based on
the fermionic projective ansatz. The Gutzwiller-projected wave
function in the $J_{1}-J_{2}$ model for $0\leq J_{2}\leq1$ is given
by
\begin{eqnarray}
|\Psi_{AF-fRVB}\rangle=\hat{P}_{J}\hat{P}_{G}|\Psi_{0}\rangle,\label{e:eq02}
\end{eqnarray}
where
$\hat{P}_{G}\equiv\prod_{i}\left(1-\hat{n}_{i\uparrow}\hat{n}_{i\downarrow}\right)$
and $\hat{P}_{J}$ is the spin-spin Jastrow correlator.
$\hat{n}_{i\sigma}=c_{i\sigma}^{\dag}c_{i\sigma}$ is the local
density. The mean-field wave function $|\Psi_{0}\rangle$ is
constructed by diagonalizing the mean-field Hamiltonian,
\begin{eqnarray}
H_{MF}=&-&\sum_{<i,j>,\sigma}t_{ij}c_{i\sigma}^{\dag}c_{j\sigma}+\sum_{i,j}\Delta_{ij}c_{i\uparrow}^{\dag}c_{i\downarrow}^{\dag}+H.c.\nonumber\\
&+&\sum_{i,\sigma}\sigma
m_{i}c_{i\sigma}^{\dag}c_{i\sigma}.\label{e:eq03}
\end{eqnarray}
Here we only consider the nearest neighbor hopping and
$t_{ij}\equiv1$. The real pairing amplitude $\Delta_{ij}$ is taken
as the nearest neighbor ($\Delta_{1}$) term and the next nearest neighbor
($\Delta_{2}$) term. They can have different values along directions
mutually perpendicular, e.g. $\Delta_{1,x},\Delta_{1,y}$
and $\Delta_{2,x+y},\Delta_{2,x-y}$.

Based on the pairing symmetry, the projected state with the
constraint of one fermion per site can describe different spin
liquids with the gapped or gapless spinon spectrum \cite{WenPRB02}.
Depending on the sign structure of $\Delta_{1}$ and $\Delta_{2}$
along both perpendicular directions, we can have different pairing
symmetries for $\Delta_{1}$ and $\Delta_{2}$: (1) $d_{x^{2}-y^{2}}$
($\Delta_{1,x}\Delta_{1,y}<0$) and $s^{++}$
($\Delta_{1,x}\Delta_{1,y}>0$) for $\Delta_{1}$; (2) $d_{xy}$
($\Delta_{2,x+y}\Delta_{2,x-y}<0$) and $s^{+-}$
($\Delta_{2,x+y}\Delta_{2,x-y}>0$) for $\Delta_{2}$. Notations of
$s^{++}$ and $s^{+-}$ are often used in the field of iron-based
superconductors \cite{HirschfeldRPP11}. The AF order parameter
$m_{i}$ can have two spatial patterns: checkerboard ($J_{2}=0$) or
collinear ($J_{2}=1$). Once the pattern is given, the amplitude of
the order parameter would be homogeneous at each site, namely,
$|m_{i}|\equiv m$. The variational degree of freedom from the AF
order can help capture the exact phase diagram in contrast to the
simple fermionic RVB states.

The spin-spin Jastrow correlator $\hat{P}_{J}$ is defined as
\begin{eqnarray}
e^{\sum_{i<j}\kappa_{ij}\hat{S}_{i}^{z}\hat{S}_{j}^{z}},\label{e:eq04}
\end{eqnarray}
where
$\kappa_{ij}\equiv\ln(r_{ij}^{\beta}w_{\gamma}^{\delta_{j,i+\gamma}})$.
Here $r_{ij}$ is the chord length of $|\vec{r}_{i}-\vec{r}_{j}|$ and
$\hat{S}_{i}^{z}$ is the spin operator along the $z$-direction at
site $i$. In addition to the parameter $\beta$ controlling
long-range spin correlations, we further consider the other three
parameters $w_{\gamma}$ for the nearest ($\gamma=1$), second-nearest
($\gamma=2$) and third-nearest ($\gamma=3$) neighbor spin-spin
correlation. The Jastrow correlator $\hat{P}_{J}$ can describe the
ferromagnetic (antiferromagnetic) correlation if
$S_{i}^{z}S_{j}^{z}>0$ ($<0$). In the case of $w_{\gamma}<1$, for
example, the short-range ferromagnetic (antiferromagnetic)
correlation would be suppressed (enhanced). On the other hand, the
factor $r_{ij}^{\beta}$ controls the short-range ($r_{ij}<1$) and
long-range ($r_{ij}>1$) correlation in an opposite way. In the
long-range case, for instance, it would decrease (increase)
ferromagnetic (antiferromagnetic) correlation if $\beta<0$. In the
following, we would illustrate that only seven variational
parameters are needed to optimize energy, which are
$\Delta,\Delta',m,w_{\gamma=1,2,3},\beta$, and almost reach the same
energy as the tensor-network state with a large number of
variational parameters.

\begin{figure}[t]
\includegraphics[height=3.4in,width=3.2in]{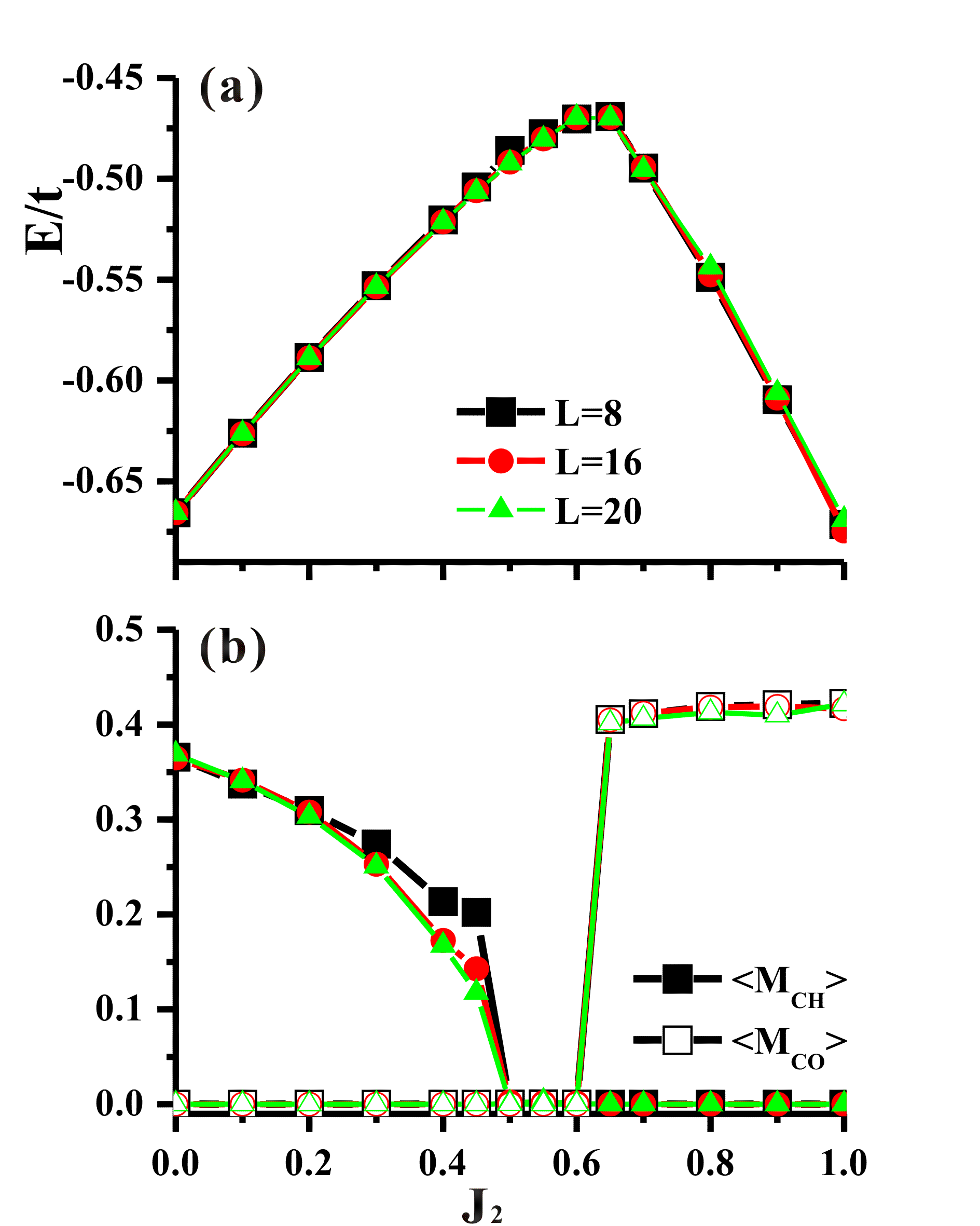}
\caption{(a) The optimized energy per site of the AF-fRVB states on
the lattice with $N=L\times L$ sites versus $J_{2}$. (b) The
variational phase diagram for the $J_{1}-J_{2}$ model on the square
lattice. Filled (Empty) symbols represent the checkerboard
(collinear) AF magnetization $\langle M_{CH}\rangle$ ($\langle
M_{CO}\rangle$). Different colors denote different lattice size as
shown in (a).}\label{fig1}
\end{figure}

\textit{Results.} Fig. \ref{fig1}(a) reveals the optimized energy of
the AF-fRVB state with both RVB correlations and magnetic orders.
The fermionic ansatz for the ground-state wave function including
the RVB pairing and the long-range AF order successfully reproduces
the frustration-induced maximum of the ground-state energy versus
$J_{2}$ obtained by several ED results
\cite{DagottoRL89,SchulzJI96,RichterEPJB10}. The optimized energy
shows much weaker size dependence than the magnetization in the
intermediate regime ($0.3\leq J_{2}\leq0.5$), as shown in
Fig.\ref{fig1}(b). For two extreme cases: $J_{2}\sim0$ and
$J_{2}\sim1$, the AF-fRVB wave function can approach the AF state
associated with the checkerboard or collinear pattern.

Two interesting phenomena should be emphasized. First, the finite
size calculation shows that the checkerboard AF phase would survive
from $J_{2}=0$ to $0.5$. At a first glance this result seems to be
inconsistent with the ground-state phase diagram obtained by ED and
DMRG. However, in the regime of $0.3\leq J_{2}\leq0.5$ the
magnetization $\langle M_{CH}\rangle$ is obviously reduced by
increasing the size of a lattice. It is necessary to conclude the
position of the transition point by further examining the larger
lattice size. In addition, the collinear AF phase suddenly appears
at $J_{2}=0.6$, implying a possible first-order transition.
Secondly, the AF-fRVB wave function would go back to the fermionic
RVB state without any magnetic order in the highly frustrated
regime, $0.5<J_{2}<0.6$, which displays the typical behavior of the
quantum spin liquid.

\begin{figure}[t]
\includegraphics[height=1.5in,width=3.4in]{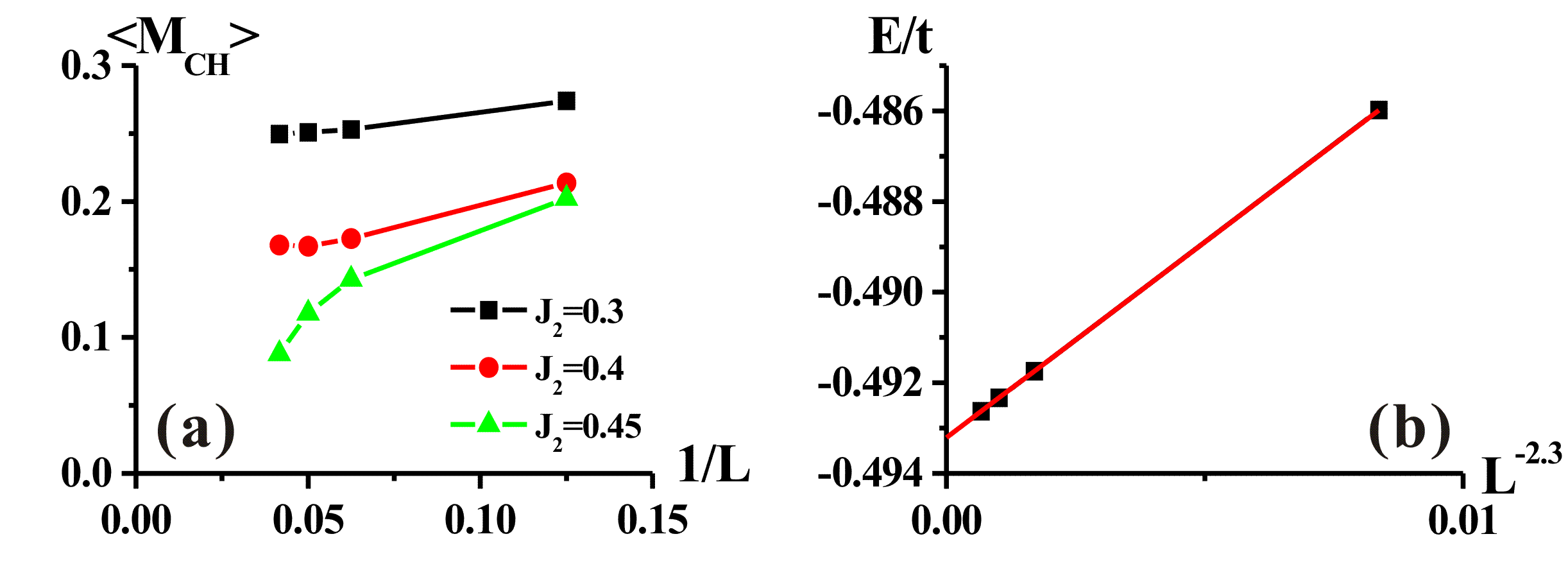}
\caption{Finite-size scaling of (a) the checkerboard AF
magnetization $\langle M_{CH}\rangle$ for different $J_{2}$ and (b)
the variational ground-state energy at $J_{2}=0.5$.}\label{fig2}
\end{figure}

In order to examine the phase transition point, we calculate the
finite size scaling of the magnetization. In Fig.\ref{fig2}(a), it
is obvious that the checkerboard AF magnetization $\langle
M_{CH}\rangle$ approaches zero at $J_{2}=0.45$ in the thermodynamic
limit, in contrast to cases of $J_{2}=0.3$ and $0.4$. Thus the
transition point between the checkerboard AF state and the spin liquid
can be clearly estimated around $J_{2}=0.45$ which is closer to
recent DMRG results \cite{GongArXiv13}. On the other hand, at the
strongest frustration point ($J_{2}=0.5$), Fig.\ref{fig2}(b) shows
that the ground-state energy per site is extrapolated to
$-0.4932(1)$ by using size scaling with the finite exponent, $-2.3$.
The ground-state energy of the AF-fRVB state only with seven
parameters is very close to $-0.4943(7)$ obtained by the PEPS with
rather large bond dimension in the thermodynamic limit
\cite{WangArXiv}, and also much lower than $-0.4893(2)$ acquired by the
Schwinger bosonic RVB wave function \cite{QiArXiv}. Moreover, the
AF-fRVB wave function on a $16\times16$ lattice system
surprisingly shows much lower optimized energy ($-0.4917(4)$) at
$J_{2}=0.5$ than other tensor network states, such as the
entangled-plaquette state ($-0.46299(3)$) \cite{MezzacapoPRB12} and
the renormalized tensor product state ($-0.45062$) \cite{YuPRB12}.
Therefore, the fermionic ansatz for the ground state of the
$J_{1}-J_{2}$ model not only reproduces the whole phase diagram but
also obtains a reasonable energy to further understand the behavior
of the quantum spin liquid in the intermediate regime.

\begin{figure}[t]
\includegraphics[height=3.4in,width=3.2in]{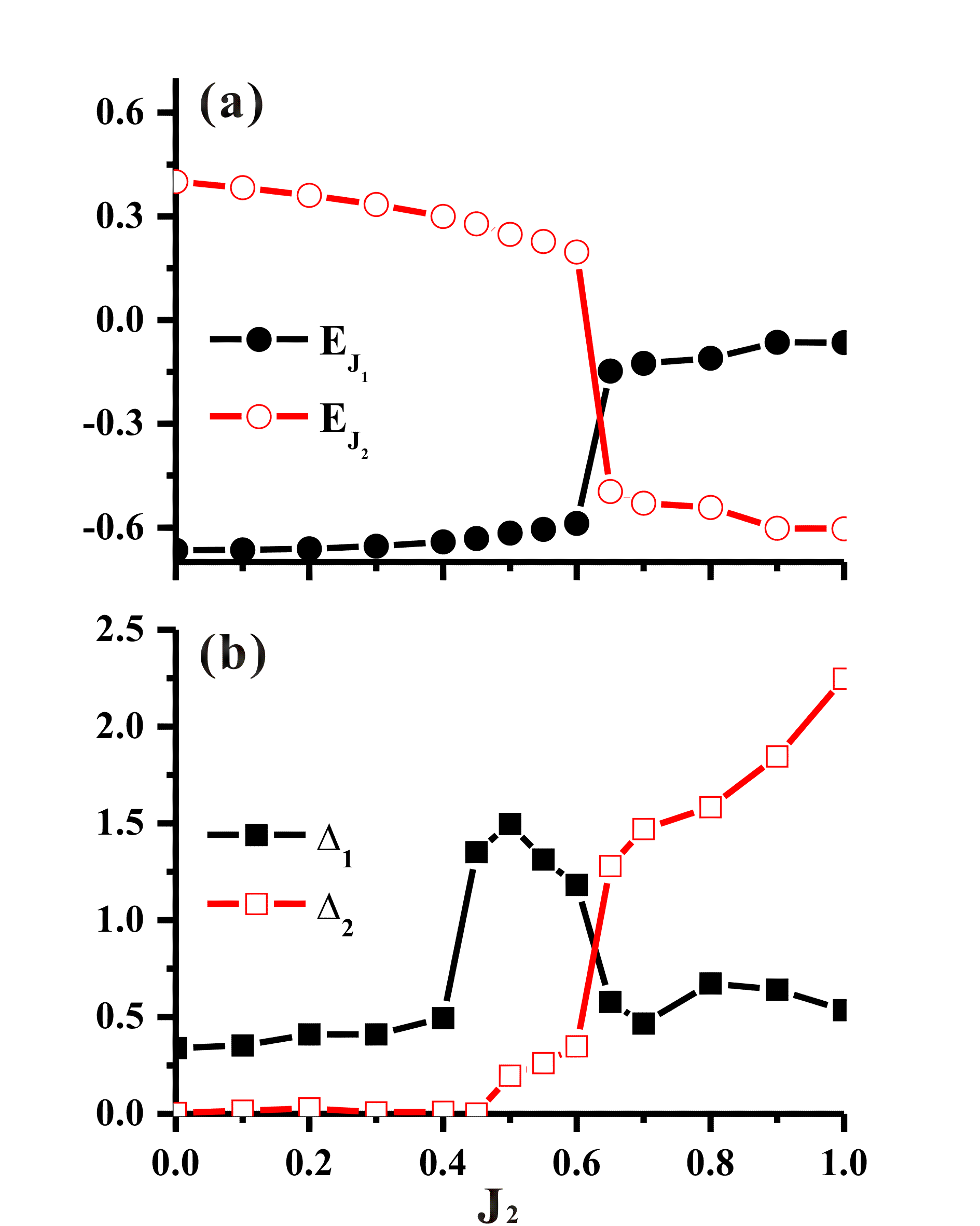}
\caption{(a) The nearest-neighbor energy $E_{J_{1}}$ and the
next-nearest-neighbor energy $E_{J_{2}}$ in the $J_{1}-J_{2}$ model
as a function of $J_{2}$. (b) The optimized parameter for the
nearest-neighbor pairing $\Delta_{1}$ and the next-nearest-neighbor
pairing $\Delta_{2}$ of the AF-fRVB state vs $J_{2}$. The size of
the lattice is $20\times20$.}\label{fig3}
\end{figure}

As pointed out in Ref.~\onlinecite{LiPRB12}, they compute the static
spin structure factor to demonstrate that a fully gapped bosonic RVB
state can capture the critical points of the $J_{1}-J_{2}$ model in
which the spin liquid is connected to the checkerboard AF phase
through a continuous transition at $J_{2}=0.4$ and to the collinear
AF state through a first-order transition at $J_{2}=0.6$. Here we
emphasize that the AF-fRVB state can easily reach qualitatively
similar conclusion as the bosonic RVB state, but is much more
accurate than the bosonic one near the magnetically ordered regime.
In Fig.\ref{fig3}(a), either the nearest-neighbor energy
($E_{J_{1}}$) or the next-nearest-neighbor energy ($E_{J_{2}}$)
continuously changes around $J_{2}=0.45$ where the checkerboard AF
magnetization $\langle M_{CH}\rangle$ drops down to zero, thus
exhibiting a second-order transition to the spin liquid at
$J_2=0.45$. However, the energy encounter a sudden jump when the
collinear AF phase appears at $J_{2}=0.6$, which obviously indicates
a first-order phase transition.

On the other hand, variational parameters about the pairing of the
AF-fRVB state also display the peculiar behavior as these magnetic
phases are transited to the spin-liquid state. In Fig.\ref{fig3}(b),
we show that the pairing parameter $\Delta_{2}$ with $d_{xy}$
symmetry is nonzero in the regime $J_{2}\gtrsim0.45$ and rapidly
increased beyond $J_{2}=0.6$. More explicitly, the nonvanishing
$\Delta_{2}$ of the AF-fRVB wave function breaks the $U(1)$ gauge
symmetry, and makes the $Z_{2}$ symmetry for the quantum spin
liquid. Furthermore, the pairing $\Delta_{1}$ with $d_{x^{2}-y^{2}}$
symmetry shows a bump in the spin-liquid regime ($0.45\leq
J_{2}\leq0.6$) which would favor to stabilize the spin-liquid phase.
Together with the vanishing magnetization, therefore, these two
optimized pairings $\Delta_{1}$ and $\Delta_{2}$ of the AF-fRVB
state give evidence for the existence of quantum spin liquid in the
frustrated regime of the $J_{1}-J_{2}$ model.

\begin{figure}[t]
\includegraphics[height=3.3in,width=3in]{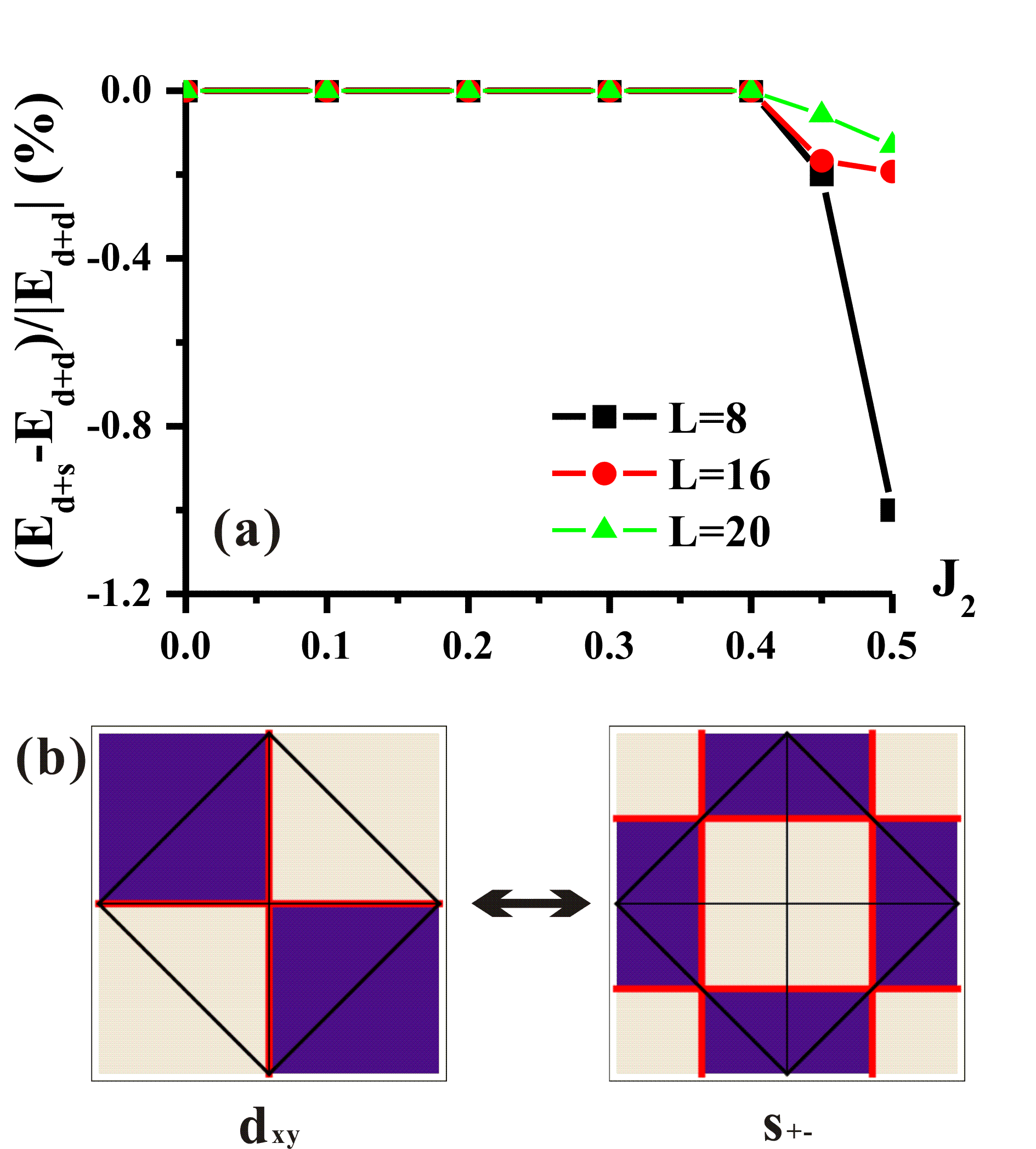}
\caption{(a) The percentage of the energy difference between
$d_{x^{2}-y^{2}}+d_{xy}$ ($d+d$) and $d_{x^{2}-y^{2}}+s_{+-}$
($d+s$) AF-fRVB states as a function of $J_{2}$ for different
lattice size. (b) $d_{xy}$ and $s_{+-}$ pairing structures
plotted in the first Brillouin zone. The black diamond is the spinon
Fermi surface. White (Blue) regions present the positive (negative)
pairing gap. Red lines mean the nodal line.}\label{fig4}
\end{figure}

In order to investigate possible pairing symmetries in the
spin-liquid phase of the frustrated model, we also consider
different pairing structures of the AF-fRVB state. Due to the
pattern of the local Hamiltonian, we examine both
$s_{++}$/$d_{x^{2}-y^{2}}$ symmetry for the nearest-neighbor pairing
$\Delta_{1}$ and $s_{+-}$/$d_{xy}$ for the next-nearest-neighbor
$\Delta_{2}$. It is worth pointing out that the variational energy
cannot be optimized if we consider $s_{++}$ symmetry in $\Delta_{1}$
(not shown). The reason is that the $d_{x^{2}-y^{2}}$ singlet pair
would avoid the on-site Coulomb repulsion so that the
$d_{x^{2}-y^{2}}$ symmetry is more favorable than $s_{++}$ in the
Heisenberg model \cite{ScalapinoPR95}. As for $\Delta_{2}$,
interestingly, the $s_{+-}$ symmetry always competes with the
$d_{xy}$ symmetry for $J_{2}\leq0.4$ as shown in Fig.\ref{fig4}(a).

In the intermediate regime ($0.4<J_{2}\leq0.5$), nevertheless, the
AF-fRVB wave function with $s_{+-}$ symmetry shows much lower energy
than the one with $d_{xy}$ symmetry in finite-size calculations.
Notably, a further finite-size analysis illustrates that the energy
difference becomes smaller as increasing the size of lattice. Thus
it is reasonable to infer that the AF-fRVB wave functions with
$s_{+-}$ and $d_{xy}$ symmetry are always degenerate in the
thermodynamic limit. Since Fig.\ref{fig2}(a) shows that there is no
long-range magnetic order ($m_{i}=0$) in the intermediate regime
($0.4<J_{2}\leq0.5$), Eq.(\ref{e:eq03}) is just the mean-field BCS
Hamiltonian that can be easily diagonalized,
\begin{eqnarray}
E_{k}=\sqrt{\varepsilon_{k}^{2}+\Delta_{k}^{2}},\label{e:mf}
\end{eqnarray}
where $\varepsilon_{k}=-2\left(\cos(k_{x})+\cos(k_{y})\right)$ and
$\Delta_{k}$ is the gap function consisting of both the
nearest-neighbor pairing $\Delta_{1}$ and the next-nearest-neighbor
pairing $\Delta_{2}$. According to the spinon spectrum $E_{k}$, the
degeneracy can be understood by their equivalent nodal structures if
we simply shift the spinon Fermi surface (black diamond,
$\varepsilon_{k}=0$) by $(\pi,\pi)$ shown in Fig.\ref{fig4}(b). Note
that plus the $d_{x^{2}-y^{2}}$-wave form factor, the energy
dispersion $E_{k}$ with both $d_{x^{2}-y^{2}}$ and $s_{+-}$ pairing
symmetries ($d+s$) clearly displays four nodes at
$(\pm\pi/2,\pm\pi/2)$. Therefore, the spin-liquid state should be
gapless because of the mean-field spinon spectrum with four Dirac
points at $(\pm\pi/2,\pm\pi/2)$ for the $d+s$ pairing structure.

\textit{Conclusions.} We have numerically studied the ground-state
phase diagram of the $J_{1}-J_{2}$ Heisenberg model on a square
lattice based on the AF-fRVB wave function. The fermionic ansatz
with long-range AF orders has successfully reproduced the
ground-state phase diagram from ED \cite{RichterEPJB10} and DMRG
\cite{JiangPRB12} calculations. The AF-fRVB wave function also
captures a second-order transition at $J_{2}=0.45$ and a first-order
transition at $J_{2}=0.6$ that is consistent with the conclusion
made by the bosonic RVB states \cite{LiPRB12}. The AF-fRVB state
naturally solves the problem that the purely fermionic RVB state
cannot describe the magnetic ordered state for $J_{2}<0.45$ and
$J_{2}>0.6$. We have shown that the AF-fRVB wave function with few
variational parameters can reach almost the same energy as the PEPS
with very large bond dimension. In addition, although the mean-field
spinon spectrum is gapful for $d_{x^{2}-y^{2}}+d_{xy}$ symmetry, the
degeneracy from $s_{+-}$ and $d_{xy}$ pairing symmetries in the
frustrated regime suggests that the spin-liquid phase can also
exhibit Dirac spinon spectrum as a result of $d+s$ pairing symmetry.

We would like to thank F. Yang and T. Ma for useful discussions.
C.P.C. is supported by Chinese Academy of Engineering Physics and
Ministry of Science and Technology. The calculations are performed
in the National Center for High-performance Computing. H.Y.C. is
supported by National Science Council of Taiwan under Grant No.
NSC-101-2112-M-003-005-MY3 and National Center for Theoretical
Science of Taiwan.


\begin{references}
\bibitem{BalentsNatPhys10}L. Balents, Nat. Phys. \textbf{464}, 199 (2010).
\bibitem{NormandCP09}B. Normand, Contemp. Phys. \textbf{50}, 533 (2009).
\bibitem{MambriniPRB06}M. Mambrini, A. Lauchli, D. Poilblanc, and F. Mila, Phys. Rev. B \textbf{74}, 144422 (2006).
\bibitem{RichterEPJB10}J. Richter and J. Schulenburg, Eur. Phys. J. B \textbf{73}, 117 (2010).
\bibitem{JiangPRB12}H.-C. Jiang, H. Yao, and L. Balents, Phys. Rev. B \textbf{86}, 024424 (2012).
\bibitem{GongArXiv13}S.-S. Gong \textit{et al.}, arXiv:1311.5962 (2013).
\bibitem{CapriottiPRL01}L. Capriotti, F. Becca, A. Parola, and S. Sorella, Phys. Rev. Lett. \textbf{87}, 097201 (2001).
\bibitem{LouPRB07}J. Lou and A. W. Sandvik, Phys. Rev. B \textbf{76}, 104432 (2007).
\bibitem{MezzacapoPRB12}F. Mezzacapo, Phys. Rev. B \textbf{86}, 045115 (2012).
\bibitem{LiPRB12}T. Li, F. Becca, W. Hu, and S. Sorella, Phys. Rev. B \textbf{86}, 075111 (2012).
\bibitem{WangPRL13}L. Wang, D. Poilblanc, Z. C. Gu, X. G. Wen, and F. Verstraete, Phys. Rev. Lett. \textbf{111}, 037202 (2013).
\bibitem{HuPRB13}W. J. Hu, F. Becca, A. Parola, and S. Sorella, Phys. Rev. B \textbf{88}, 060402(R) (2013).
\bibitem{QiArXiv}Y. Qi and Z.-C. Gu, arXiv:1308.2759 (2013).
\bibitem{AndersonSci87}P. W. Anderson, Science \textbf{235}, 1196 (1987).
\bibitem{WenPRB02}X.-G. Wen, Phys. Rev. B \textbf{65}, 165113 (2002).
\bibitem{AuerbachPRL88}A. Auerbach and D. P. Arovas, Phys. Rev. Lett. \textbf{61}, 617 (1988).
\bibitem{ArovasPRB88}D. P. Arovas and A. Auerbach, Phys. Rev. B \textbf{38}, 316 (1988).
\bibitem{FlintPRB09}R. Flint and P. Coleman, Phys. Rev. B \textbf{79}, 014424 (2009).
\bibitem{YunokiPRB06}S. Yunoki and S. Sorella, Phys. Rev. B \textbf{74}, 014408 (2006).
\bibitem{RanPRL07}Y. Ran, M. Hermele, P. A. Lee, and X.-G. Wen, Phys. Rev. Lett. \textbf{98}, 117205 (2007).
\bibitem{LiArXiv11}T. Li, arXiv:1101.1352 (2011).
\bibitem{WangArXiv}L. Wang, Z.-C. Gu, F. Verstraete, and X.-G. Wen, arXiv:1112.3331 (2012).
\bibitem{DarradiPRB08}R. Darradi \textit{et al.}, Phys. Rev. B \textbf{78}, 214415 (2008).
\bibitem{CapriottiPRB03}L. Capriotti, F. Becca, A. Parola, and S. Sorella, Phys. Rev. B \textbf{67}, 212402 (2003).
\bibitem{HirschfeldRPP11}P. J. Hirschfeld, M. M. Korshunov, and I. I. Mazin, Rep. Prog. Phys. \textbf{74}, 124508 (2011).
\bibitem{DagottoRL89}E. Dagotto and A. Moreo, Phys. Rev. Lett. \textbf{63}, 2148 (1989).
\bibitem{SchulzJI96}H. J. Schulz, T. A. L. Ziman, and D. Poilblanc, J. Phys. I \textbf{6}, 675 (1996).
\bibitem{YuPRB12}J.-F. Yu and Y.-J. Kao, Phys. Rev. B \textbf{85}, 094407 (2012).
\bibitem{ScalapinoPR95}D. J. Scalapino, Phys. Rep. \textbf{250}, 329 (1995).
\end{references}
\end{document}